\documentclass[12pt]{revtex4}
\usepackage{amsmath,amssymb,bm}
\usepackage[toc,page]{appendix}
\usepackage{graphicx}
\usepackage{url}
\usepackage{hyperref,xcolor}

\DeclareGraphicsExtensions{.pdf,.png,.jpg,.eps}

\begin{document}
\title{Absence of solid phase in dense amorphous active granular matter}
\author{Cunyuan Jiang}
\affiliation{School of Physics, Zhengzhou University, Zhengzhou 450001, Henan, China}

\begin{abstract}
Solid phase of dense granular matter is inevitable because of jamming transition when the packing fraction or the pressure suffered is high enough. The experiment suggests that active Brownian granular matter will keep fluid phase even under the highest packing fraction (higher than the packing fraction of crystallization) if crystallization is prevented by mixing granular particles of different sizes. The findings encourage us to reconsider the role of activity in affecting the global dynamical properties of matter.
\end{abstract}

\maketitle
A group of hard granular particles, if they are not allowed to break each other and have no other interactions between them, for example a sand pile, will behave as a rigid solid due to jamming transition when their packing fraction and external pressure are high \cite{Deng2024,annurev:/content/journals/10.1146/annurev-conmatphys-070909-104045,PhysRevLett.84.4160,PAN20231}. On the other hand, if the packing fraction is low, the granular matter can behave as a fluid. The above picture is well established for passive granular matter, for which the collective dynamical properties are solely determined by collisions (interactions) between granular particles besides the factors of their shape. Another more common category of granular matter is active granular matter, for which each particle can independently convert energy from the environment into its own behavior \cite{Shaebani2020,Shankar2022,RevModPhys.88.045006}. Therefore, an additional important factor, activity, is introduced in active granular matter to compete with interactions between granular particles. The competition between activity and interaction will lead to unexpected collective behaviors of active granular matter \cite{annurev:/content/journals/10.1146/annurev-conmatphys-031119-050752,Chen2024}.

\begin{figure}
    \centering
    \includegraphics[width=0.5\linewidth]{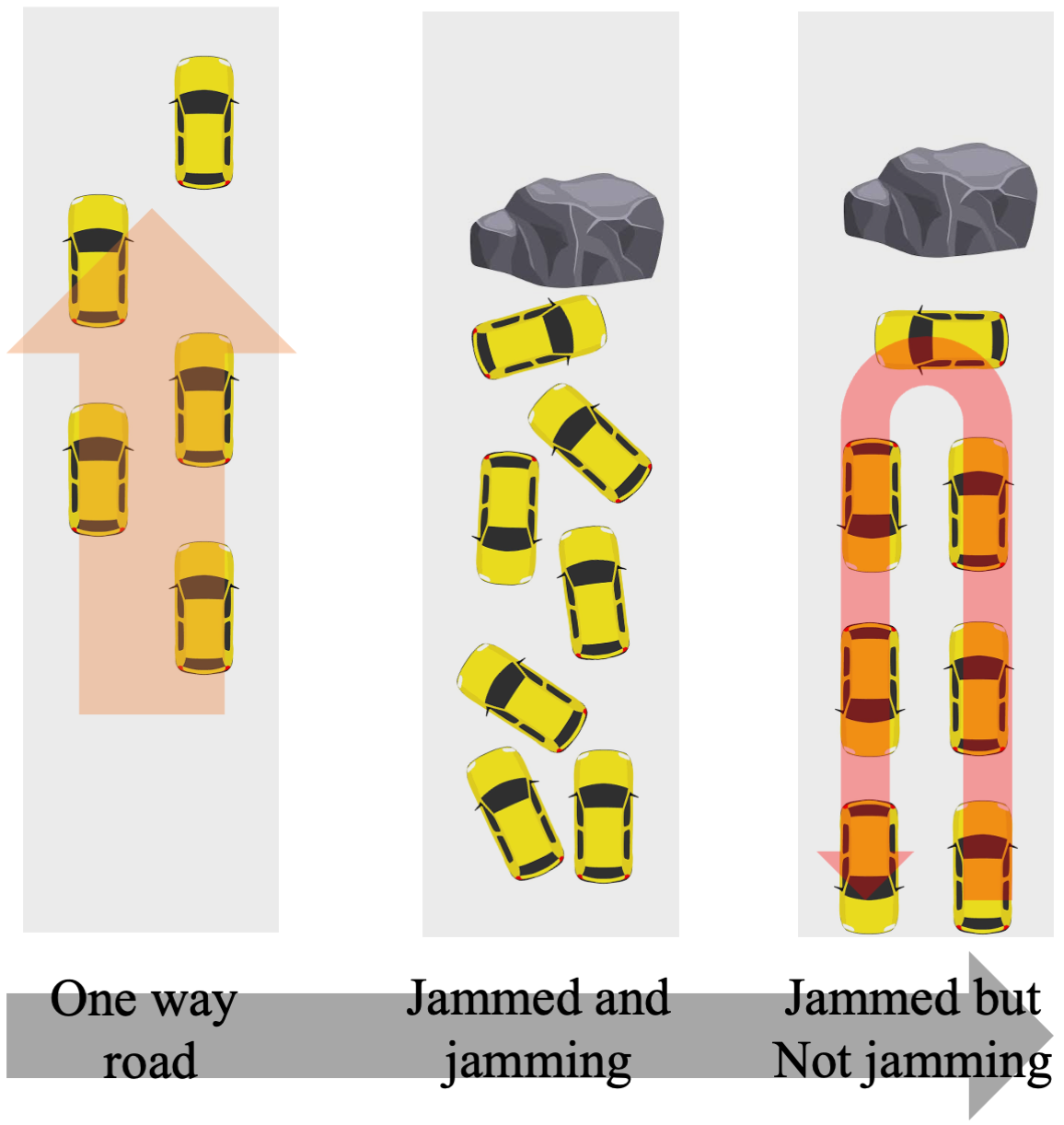}
    \caption{A normal one way highway (left), a traffic jam caused by an unexpected event (middle), and alleviation of the traffic jam (right). From left to right, the cars manifest as fluid, solid, and then fluid again when nothing happens, when an unexpected event is encountered, and sometime after that.}
    \label{figcar}
\end{figure}

A question naturally arises: what will happen to the jamming transition and solidification in active granular matter under the competition of activity and interaction? Let us consider a scenario, which inspired the author to ask this question, that many cars run on a narrow one way highway as shown on the left of Fig.\ref{figcar}. The cars in this situation belong to the phase of active granular fluid because all cars can run smoothly on the road. If an accident (for example, a rock) happens and blocks the road, the cars will behave as a solid like jamming state in a short time because cars are not allowed to turn around on the narrow one way highway, as shown in the middle of Fig.\ref{figcar}. Traffic congestion usually does not last long because drivers are active in adjusting their cars and easing the congestion to leave the road, as shown on the right of Fig.\ref{figcar}. The fluid phase of cars is thus recovered after some time following the accident. The above story of traffic jam leads us to an intuitive answer to the question about the role of activity in affecting the jamming transition of dense active granular matter: activity can prevent dense active granular matter from staying in the solid phase for a long time, and therefore the mechanism of jamming transition can only keep the dense active granular matter in the solid phase for a short period.

In addition to the above imaginative description of traffic jam, there is a work that reports the presence of collective behavior of dense human crowds during the San Fermin festival in Spain \cite{Gu2025}. They observed that about 30 minutes before the festival opening, when the density of human crowds increases to a critical value, the crowd dynamics undergoes a sharp transition captured by a linear increase of mean velocity square, augmented by activity bursts. The observed collective dynamical behaviors here contradict the conclusion for passive granular matter mentioned above, where the mean velocity square of each granular particle should be limited to zero due to the solid phase induced by jamming. The observation in Ref.\cite{Gu2025} therefore indicates the absence of solid phase in dense crowds.

However, the individuals in cars and crowds are intelligent and can control their own behaviors. What if the individuals lack autonomous decision making? Active Brownian granular particles, in which each particle undergoes a time dependent random force \cite{Romanczuk2012}, provide an ideal platform to study the question proposed at the beginning: whether activity can prevent the solid phase of granular matter under high packing fraction. In this work, we re-exam the experimental results of active Brownian particles implemented on a Brownian vibrator published on\cite{Chen2024,PhysRevE.106.L052903,PhysRevLett.133.188302,Jiang2025}. We observe that the vibrational density of states (DOS) of the experimental system remains nonzero at zero frequency under the highest packing fraction of implementation if crystallization is prevented by mixing different sizes of granular particles. This finding suggests that the solid phase cannot be reached by increasing the packing fraction and indicates the absence of jamming transition in the experimental system. The findings align with the previous discussion about traffic jam and human crowds \cite{Gu2025}, and encourage us to reconsider the role of activity in overturning our understanding of phase transition and collective dynamics of dense active granular matter.

The experimental system consists of some plastic disks (granular particles) with size on the order of one centimeter distributed on a two dimensional horizontal Brownian vibrator. The space (and also the boundary of the vibrator platform) that allows particles to move is fixed. The ratio of the diameter of big and small particles is \(1.4:1\) and the mixing ratio is \(1:2\) for the bi-dispersed situation and \(0:1\) for the crystal situation. The driving frequency of the Brownian vibrator is fixed to \(100\) Hz. During the experiments, the positions of each particle are recorded by a camera at a rate of \(40\) frames per second. The readers are encouraged to refer to Refs.\cite{Chen2024,PhysRevE.106.L052903,PhysRevLett.133.188302,Jiang2025} for more details about the experimental setup. Once the trajectories of particles are known, the dynamical properties of the active Brownian particles as a whole can be obtained through the displacement correlation matrix \(C_{ij} = \langle d(t)_i d(t)_j \rangle_t\) and the dynamical matrix \(D_{ij} \propto (C^{-1})_{ij} / \sqrt{m_i m_j}\), where indices \(i(j)\) refer to the degrees of freedom (DOFs) of the system, \(\langle \rangle_t\) denotes time average, \(d(t)_{i(j)}\) denotes the displacement from the average position of the \(i(j)\)th DOF, and \(m_{i(j)}\) is the mass of the particle \cite{PhysRevLett.105.025501}. The DOS can be obtained by the statistics of the distribution of eigenfrequencies of the system, which are the square roots of the eigenvalues of the dynamical matrix.

DOS can be used to identify whether the phase of matter is solid or liquid. For a solid phase, the collective dynamical behaviors should be sound waves, and therefore the DOS should satisfy Debye's theory, which predicts linear DOS as a function of frequency, \(\propto \omega\), in the low frequency region for a two dimensional system \cite{debye}. For a liquid phase, there should be additional low frequency DOS, especially nonzero DOS at zero frequency due to the flow of liquid \cite{Jiang2025}.

\begin{figure}
    \centering
    \includegraphics[width=0.5\linewidth]{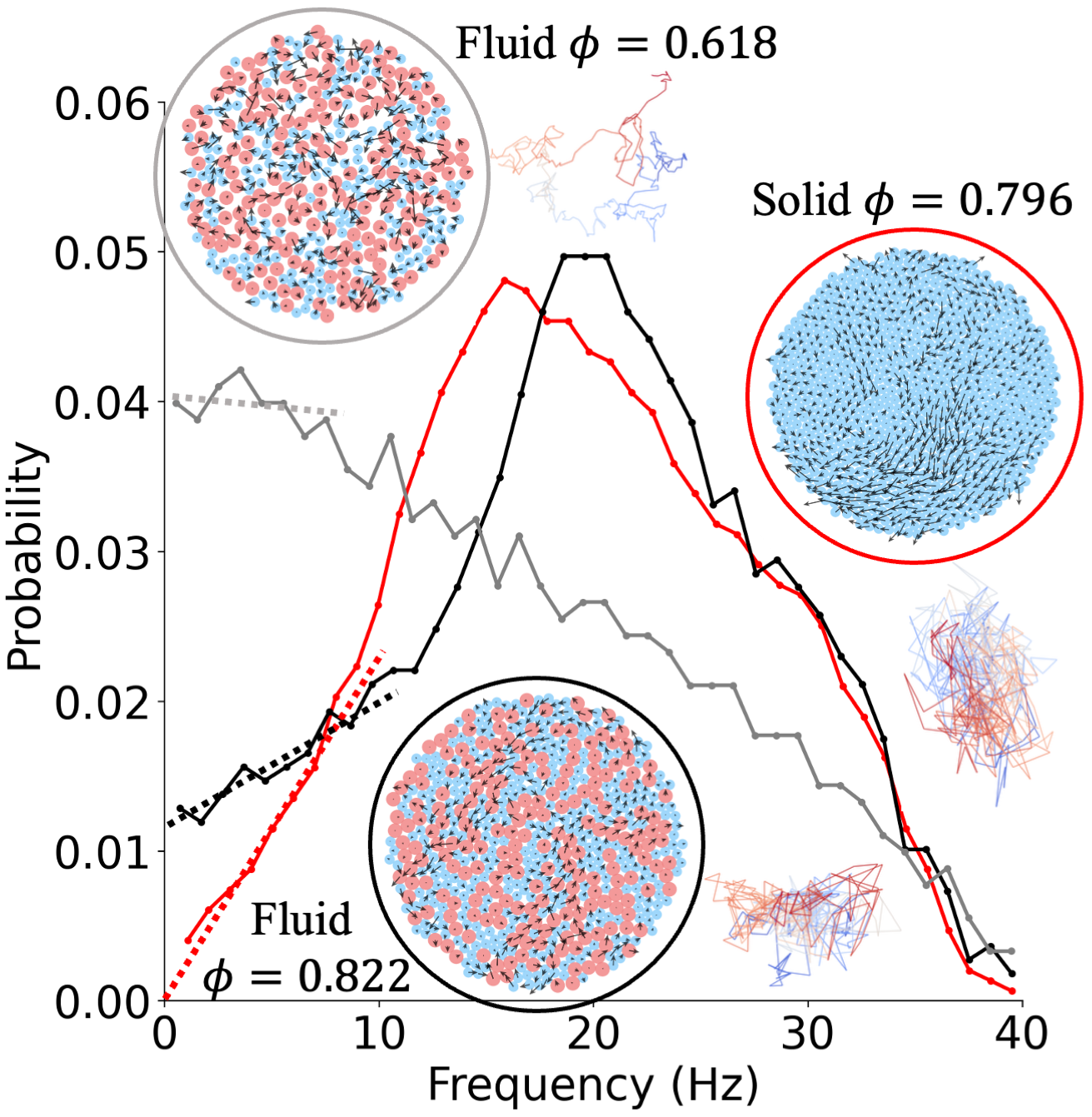}
    \caption{The VDOS (probability of eigenfrequencies as a function of frequency) of active Brownian granular matter with bi-dispersed packing fraction \(\phi = 0.618\) (gray), bi-dispersed packing fraction \(\phi = 0.822\) (black), and mono dispersed packing fraction \(\phi = 0.796\) (red). The inserts in circles with corresponding colors show their structure and typical displacement field under observation time \(10\) s. Big and small particles are distinguished by red and blue respectively. Typical trajectories of a single particle for each situation are shown associated with their displacement field. The time of the trajectories is distinguished by color from blue to red as time goes by.}
    \label{figdos}
\end{figure}

In Fig.\ref{figdos}, the DOS of three situations are shown: low packing fraction \(\phi = 0.618\) bi-dispersed (gray), high packing fraction \(\phi = 0.796\) mono dispersed (red), and high packing fraction \(\phi = 0.822\) bi-dispersed (black). The structure and typical trajectory of a particle for the three cases are shown as insets of Fig.\ref{figdos}. According to the previous criterion of phase, the two cases, \(\phi = 0.618\) bi-dispersed and \(\phi = 0.822\) bi-dispersed, should be classified as fluid phase, while the case \(\phi = 0.796\) mono dispersed should be classified as solid phase. It is not surprising that the case \(\phi = 0.618\) bi-dispersed is fluid phase because the packing fraction is not high enough to trigger jamming transition and turn the dynamics into behaving as a solid. It is also not surprising that the case \(\phi = 0.796\) mono dispersed belongs to solid phase because of crystallization. Crystallization is the destiny of symmetry when the packing fraction of a mono dispersed system is high enough. Under a crystal phase with discrete translational symmetry, sound waves as spontaneously broken translational symmetry are the only allowed low energy and hence low frequency excitations. Therefore, the DOS of the crystal phase must agree with the prediction of Debye's theory, which gives linear DOS at low frequency as shown in Fig.\ref{figdos} in red. The experimental results of the \(\phi = 0.796\) mono dispersed case agree with Debye's law and hence should be classified as solid phase. Its DOS peaks at \(15\) Hz, which is a Van Hove peak due to interference between sound waves and the crystal lattice.

The DOS of the \(\phi = 0.822\) bi-dispersed case is special because of two features as shown in Fig.\ref{figdos} in black. The first feature is that the DOS at zero frequency is nonzero, which indicates liquid phase at low frequency. The second feature is that there is still a Van Hove peak at high frequency around \(20\) Hz, which indicates solid phase at high frequency. The frequency of the Van Hove peak for the \(\phi = 0.822\) bi-dispersed case is higher than that for the \(\phi = 0.796\) mono dispersed case due to higher packing fraction and hence harder material and larger speed of sound. Frequency is the inverse of time scale; the \(\phi = 0.822\) bi-dispersed case behaves as a solid under short time dynamical response but as a fluid under long time. Therefore the case of \(\phi = 0.822\) bi-dispersed is in fact a kind of viscous liquid. A point that must be noticed is the packing fraction here is even higher than that of the crystal phase. According to the jamming phase diagram of granular matter \cite{annurev:/content/journals/10.1146/annurev-conmatphys-070909-104045}, a higher packing fraction leads to a more solid like overall dynamics. The observation of the \(\phi = 0.822\) bi-dispersed case violates the conclusion from general passive granular matter and indicates the absence of solid phase in such active Brownian particles if crystallization can be prevented.

From the \(\phi = 0.618\) bi-dispersed case to the \(\phi = 0.822\) bi-dispersed case, the system becomes more solid like due to the decreasing probability of zero frequency DOS. One may ask whether the zero frequency DOS could decrease to zero and then become completely a solid by definition if the packing fraction could be increased to a higher level. This consideration can be abandoned for two reasons. The first is that the highest packing fraction is not necessary for behaving as a solid (reducing zero frequency DOS to zero). The theoretical estimation of the highest packing fraction for mono dispersed hard disks on a two dimensional flat table is \(\phi = \pi/\sqrt{12} \sim 0.906\) (hexagonal close packing). However, as shown in Fig.\ref{figdos}, crystallization and linear DOS appear at packing fraction \(\phi = 0.796\). The second reason is that \(\phi = 0.822\) is already experimentally the highest packing fraction for the bi-dispersed situation. Experimentally, no more particles can be added to the system of \(\phi = 0.822\) bi-dispersed while keeping the boundary unchanged and without disks squeezing out of the plane. These two reasons suggest that zero frequency DOS can be reduced by increasing packing fraction but cannot be erased to zero for the bi-dispersed case.

From the behavior at different time scales, the observation of the experiments on active Brownian particles qualitatively aligns with the narrative description of traffic jam of cars at the beginning, where the jamming phase appears shortly after an accident but returns to fluid phase after a long time. From the spatial behavior, the observation from the experiments, nonzero DOS at zero frequency, also qualitatively aligns with the observation of non solid anomalous collective motion of dense human crowds in Ref.\cite{Gu2025}. All the results encourage us to draw a conclusion in a more general aspect: a pure solid phase cannot survive in dense active granular matter under long time observation. In our opinion, these observations can be interpreted as follows.

\begin{figure}
    \centering
    \includegraphics[width=0.5\linewidth]{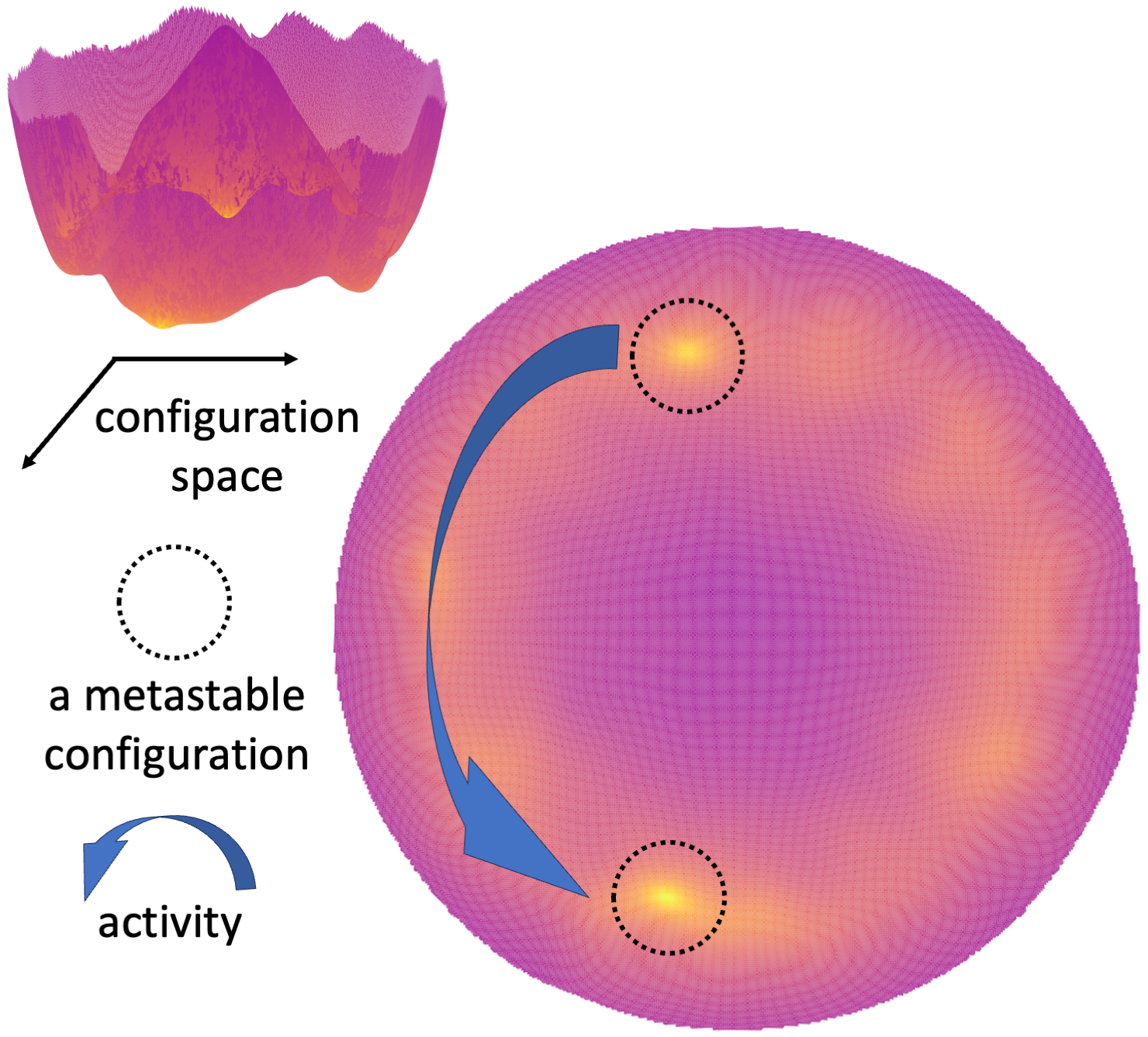}
    \caption{Schematically showing the potential of a twisted Mexican hat with many metastable minima. The color from lighter to darker indicates the energy from low to high. The blue arrow indicates the role of activity that shifts the system from one metastable state to another crossing the barrier in between. The main figure is a top view and the inset at the top left is a side view of the twisted Mexican hat.}
    \label{fighat}
\end{figure}

If crystallization is prevented, one of the most outstanding properties of an amorphous system is the presence of nearly infinite allowed (metastable) configurations under a given packing fraction due to replica symmetry breaking \cite{parisi2002physicalmeaningreplicasymmetry}. The relation between these metastable configurations can be schematically shown in Fig.\ref{fighat} as local minima of a twisted Mexican hat which is not flat at the bottom. Each minimum corresponds to a configuration of the amorphous active granular matter. Active matter is highly non equilibrium and total energy is allowed to fluctuate within a small range. The role of activity is therefore to drive the system from one configuration to another, crossing the low energy barrier between them. In real space, the change of configurations behaves as rearrangement of granular particles and hence as flows as in a fluid. Due to the huge number of metastable configurations within a very close energy range, activity can always (with very high probability) find another configuration with a small energy barrier from the current configuration. The two elements of the final picture to interpret the absence of solid phase in dense active granular matter are that disorder induces a large number of metastable configurations and activity shifts the current metastable configuration from one to another. The system travels in configuration space under activity and undergoes incessant rearrangement, behaving as a fluid.

\begin{figure}
    \centering
    \includegraphics[width=0.5\linewidth]{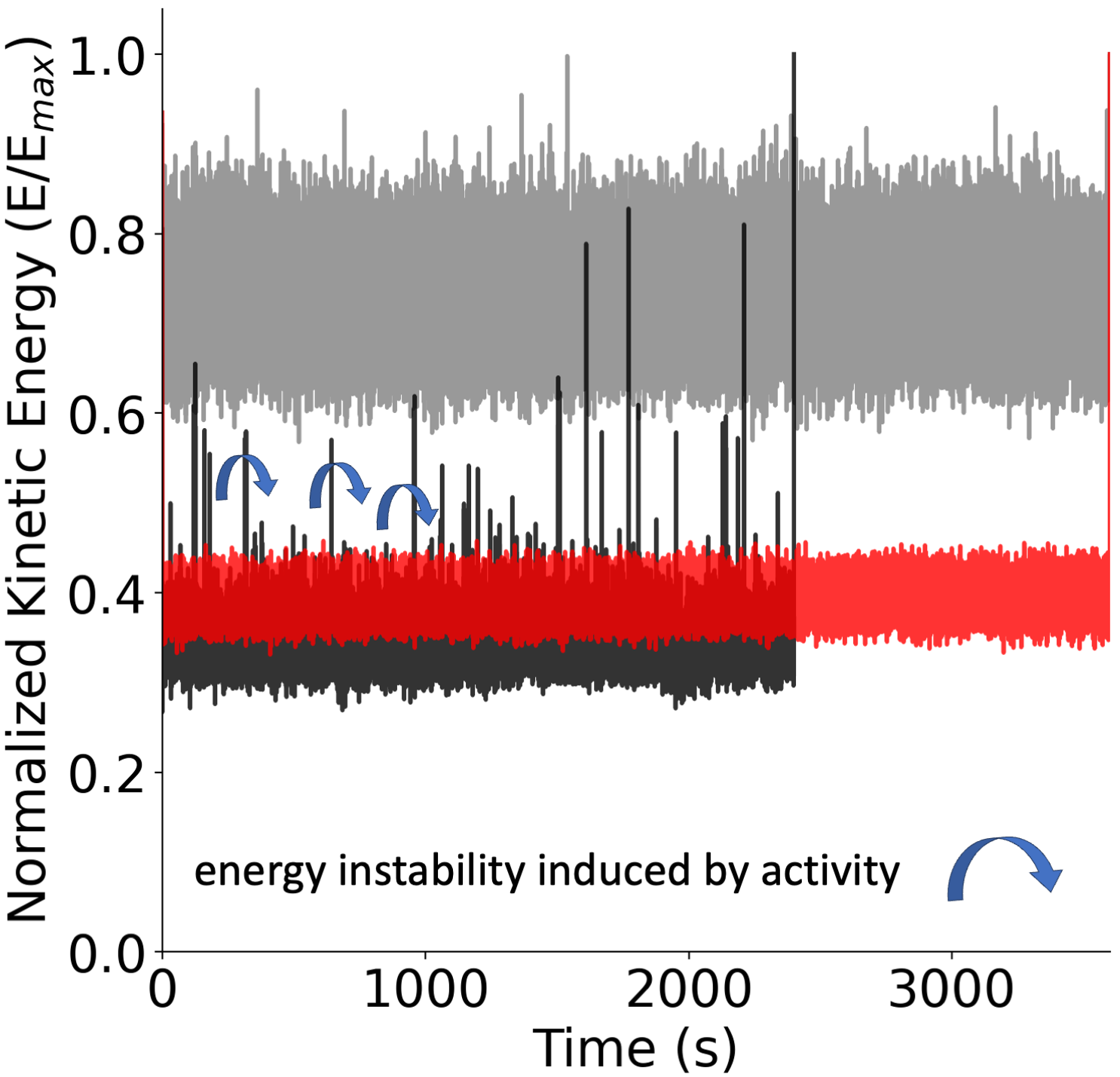}
    \caption{The total kinetic energy for different situations as a function of time. Colors gray, black, and red correspond to bi-dispersed packing fraction \(\phi = 0.618\), bi-dispersed packing fraction \(\phi = 0.822\), and mono dispersed packing fraction \(\phi = 0.796\) respectively. The sharp peaks of \(\phi = 0.822\) bi-dispersed case (black) are the jumps of total kinetic energy and indicate the system are shifting from one metastable configuration to another driven by activity.}
    \label{figenergy}
\end{figure}

We find the proof to confirm above interpretation by examining the total kinetic energy of the systems as shown in Fig.\ref{figenergy}. It can be seen that, for \(\phi = 0.796\) crystal situation, the total kinetic energy are limited to a narrow range. This result corresponds to that the crystal system vibration around a stable point of configuration space. For the dense amorphous situation, the curve of total energy have many conspicuous and sharp high energy peaks. The presence of these peaks indicates that the system undergoes a short time but high energy state, which is that the system is crossing the energy barrier between two metastable configurations as shown by Fig.\ref{fighat}. The sharp peaks are sparse comparing with the time scale. This means the jump between metastable configurations are not continuous, the system will stay at metastable configuration for some time, life time of a metastable configuration, which is around \(\sim 100 s\) in the experiment. For the situation of lower packing fraction \(\phi = 0.618\), it is a dilute fluid phase, the total kinetic energy distributed in a wider range. In this situation, there is no stable configuration of solid, and there is even no metastable configuration. It is completely random fluid configuration, for example, as shown in the insert of Fig.\ref{figdos}.

In summary, we have phenomenologically discussed the absence of jamming transition induced solid phase in cars and human crowds. The experimental results on amorphous active Brownian particles also suggest the absence of solid phase even at the highest (experimentally reachable) packing fraction. In the experiment, the liquid behavior is characterized by nonzero DOS at zero frequency. Finally, the absence of solid phase in dense amorphous active granular matter is attributed to the traveling of the system in configuration space driven by activity. Our findings provide an interpretation of the role of activity in affecting the phase of active granular matter.

The experimental data involved here have been published in Ref.\cite{Chen2024,PhysRevE.106.L052903,PhysRevLett.133.188302,Jiang2025}. The opinion of this work is independent from the authors of publications.

The author would like to thank Yangrui Chen for enlightening discussions.


\end{document}